\documentclass[conference]{IEEEtran}
\IEEEoverridecommandlockouts
\usepackage{cite}
\usepackage{amsmath,amssymb,amsfonts}
\usepackage{algorithmic}
\usepackage{graphicx}
\usepackage{textcomp}
\usepackage{xcolor}
\usepackage{booktabs}
\usepackage{multirow} 
\usepackage{cite} 
\usepackage[numbers,compress]{natbib}
\makeatletter
\newcommand{\linebreakand}{%
  \end{@IEEEauthorhalign}
  \hfill\mbox{}\par
  \mbox{}\hfill\begin{@IEEEauthorhalign}
}
\makeatother

\def\BibTeX{{\rm B\kern-.05em{\sc i\kern-.025em b}\kern-.08em
    T\kern-.1667em\lower.7ex\hbox{E}\kern-.125emX}}
\begin{document}

\title{Fragile Model Watermark for integrity protection: leveraging boundary volatility and sensitive sample-pairing\\
{
}
\thanks{The work is supported by National Natural Science Foundation of China No.62172001 and Research Fund of Guangxi Key Lab of Multi-source Information Mining \& Security (MIMS23-M-04).}

}
\author{\IEEEauthorblockN{1\textsuperscript{st} ZhenZhe Gao}
\IEEEauthorblockA{\textit{Shanghai Key Laboratory of } \\
\textit{Multidimensional Information Processing,}\\
\textit{East China Normal University}\\
Shanghai, China \\
zzgao@stu.ecnu.edu.cn}
\and
\IEEEauthorblockN{2\textsuperscript{nd} Zhenjun Tang}
\IEEEauthorblockA{\textit{Guangxi Key Lab of Multi-source } \\
\textit{Information Mining \& Security,}\\
\textit{Guangxi Normal University}\\
Guilin, China \\
zjtang@gxnu.edu.cn}
\and
\IEEEauthorblockN{3\textsuperscript{rd} Zhaoxia Yin*}
\IEEEauthorblockA{\textit{Shanghai Key Laboratory of } \\
\textit{Multidimensional Information Processing,}\\
\textit{East China Normal University}\\
Shanghai, China \\
zxyin@cee.ecnu.edu.cn}
\linebreakand
\IEEEauthorblockN{4\textsuperscript{th} Baoyuan Wu}
\IEEEauthorblockA{\textit{Shenzhen Research Institute of Big Data,} \\
\textit{The Chinese University of Hong Kong}\\
Shenzhen, China \\
wubaoyuan@cuhk.edu.cn}
\and
\IEEEauthorblockN{5\textsuperscript{th} Yue Lu}
\IEEEauthorblockA{\textit{Shanghai Key Laboratory of } \\
\textit{Multidimensional Information Processing,}\\
\textit{East China Normal University}\\
Shanghai, China \\
ylu@cs.ecnu.edu.cn}}

\maketitle

\begin{abstract}
Neural networks have increasingly influenced people's lives. Ensuring the faithful deployment of neural networks as designed by their model owners is crucial, as they may be susceptible to various malicious or unintentional modifications, such as backdooring and poisoning attacks.
Fragile model watermarks aim to prevent unexpected tampering that could lead DNN models to make incorrect decisions. They ensure the detection of any tampering with the model as sensitively as possible.
However, prior watermarking methods suffered from inefficient sample generation and insufficient sensitivity, limiting their practical applicability.
Our approach employs a sample-pairing technique, placing the model boundaries between pairs of samples, while simultaneously maximizing logits. This ensures that the model's decision results of sensitive samples change as much as possible and the Top-1 labels easily alter regardless of the direction it moves. 
Experimental evaluations conducted across multiple models and datasets demonstrate the superior sensitivity and generation efficiency of our method compared to the current approaches.
\end{abstract}


\section{Introduction}
Deep Neural Networks (DNNs) have been widely applied across various domains. However, this has also sparked concerns regarding the trustworthiness and safety of artificial intelligence, encompassing issues such as model integrity preservation and intellectual property (IP) rights protection. 

Ensuring model integrity aims to prevent intentional or unintentional tampering that could lead users to make incorrect decisions. This is distinct from robust watermarks used for model IP  protection \cite{ICME1}.  
Fragile watermarks emphasize safeguarding sensitivity, ensuring the detection of any tampering to the model as sensitively as possible \cite{sensitive}.

Fragile watermarking can counter the risk that the deployed model could undergo various modifications, which may potentially compromising its intended functionality \cite{Trojan}.
Modification might involve fine-tuning the model to embed backdoor trigger samples \cite{badnets}, making significant adjustments to model parameters, or pruning \cite{prune} and quantizing \cite{quantizing,quantizing2} the model to reduce computational and storage resources required by cloud providers.
However, once a model is deployed, whether locally or through cloud-based Machine Learning as a Service (MLaaS) \cite{mlaas}, users typically lack the permission to access its internal parameter structure. In this context, detecting the state of a model often only captures the model's output results. This type of detection, which doesn't require examining the internal parameters of the model, is evidently more universal.

\begin{figure}[tb]
	\centering
		\includegraphics[scale=0.8]{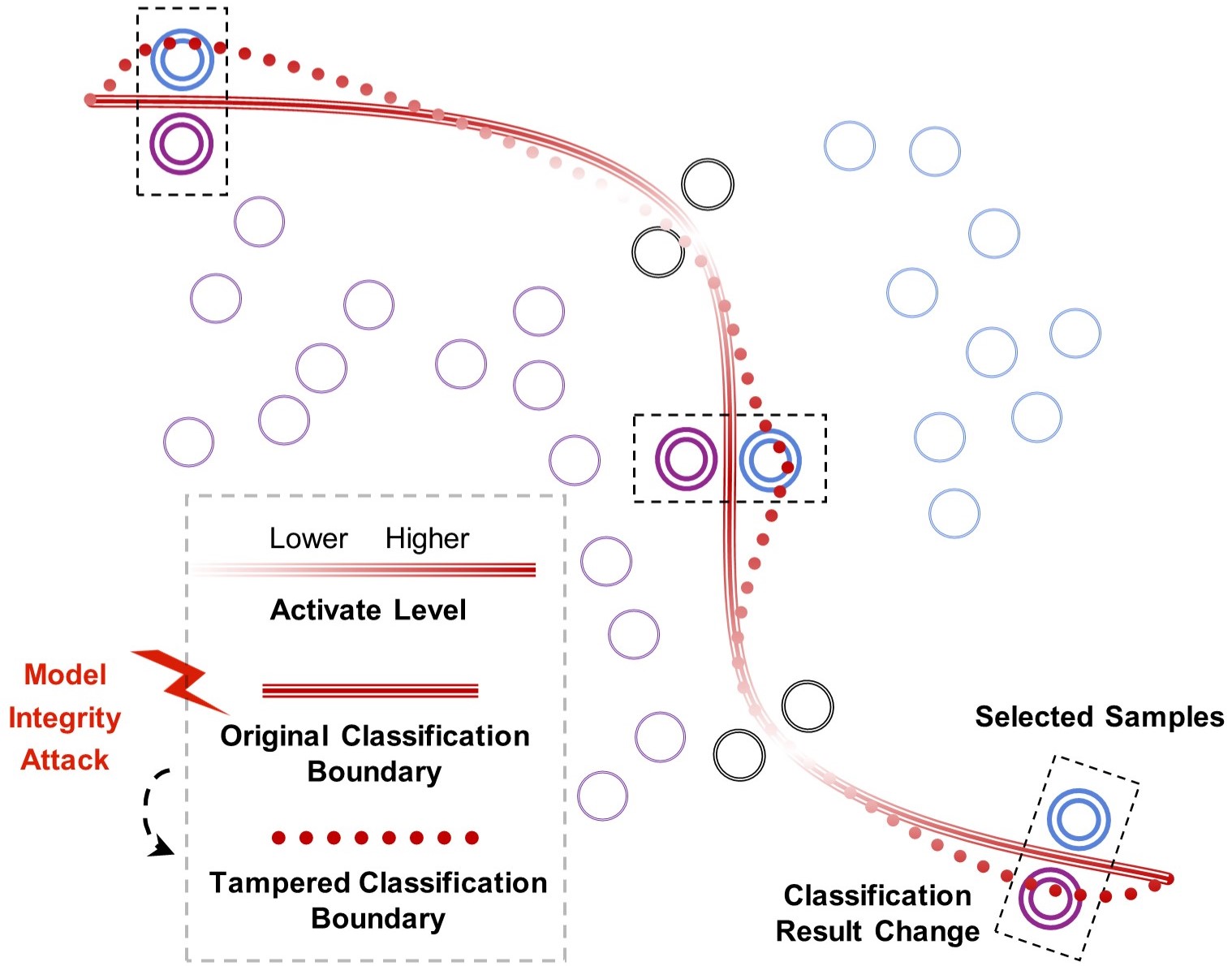}
	\caption{\small Demonstration of model boundary alterations and positions of sample pairs selected by using the proposed method. The circles represent different samples. Here, `activate level' indicates the degree of activation logits of the samples near the model boundary, a concept further elaborated in Section 3.B.}
	\label{figure1}
\end{figure}

Previous researchers like He et al.\cite{sensitive}, considered protecting neural networks by amplifying the gradient of the network's output with respect to certain preset parameters when sensitive samples are input without the need to inspect internal parameters to detect the model's state. However, when the model is modified, it is difficult for us to know a priori which part of the parameters the adversary will modify. Aramoon et al.\cite{aid} aimed for sensitive samples to activate a greater number of neurons, thereby safeguarding the entire network. However, achieving these objectives becomes challenging in scenarios involving a large number of classifications (greater than 10) and deep network structures, where the complexity of neuronal interactions increases, making it difficult to traverse to activate all neurons. 
Other researchers, such as Yin et al.\cite{yin2023decision}, generated sensitive samples using methods similar to adversarial attacks \cite{chen2023advfas,cheng2023topology,tang2024wfss} to approach the model boundary. However, such methods, due to the lack of deliberate steps to activate neurons, carry the possibility that although the sensitive samples are located at the model boundary but does not change significantly.

In this paper, building upon considerations of prior work, we believe that model integrity safety needs attention in the following areas: (1) Carefully crafted sensitive samples should be easily producible without consuming excessive resources.
(2) The sensitivity of generated samples should be maximized to detect changes in any model adjustments.
To address these concerns, we propose a method for generating sensitive samples by analyzing the boundaries of classification models. Specifically, in the context of multi-class classification, we tackle the challenges of sensitive sample convergence by incorporating one additional binary classification layer. This layer serves to simplify the relationship between sensitive samples and model boundaries. Furthermore, we activate neurons by utilizing output logits exclusively, avoiding the need to iterate through all neurons and thus mitigating the substantial computational cost.

In our approach, we emphasize two key elements necessary for sensitive samples to effectively perceive the movement of model boundaries: (a) Post-adjustment of model parameters, the output variations in response to sensitive samples should be maximized, which means that sensitive samples should be positioned in an area where minor model adjustments ($\Delta x$) lead to the largest possible change in model output ($\Delta y$). (b) Given the unpredictability of the direction in which model boundaries may shift, it is optimal for sensitive samples to be positioned on both sides of the model boundary. The location demonstration of the sensitive samples we generated is shown in Figure \ref{figure1}.

The contributions of this paper are:

\begin{itemize}

    \item To enhance sensitivity, we analyze the characteristics of model boundaries and introduce a loss function to approach the most volatile boundaries regions.
    \item To further enhance sensitivity, we use a two-stage sample generation process to generate sample pairs so that the model boundary is sandwiched between the sample pairs. 
    \item To enhance efficiency, we employ one additional binary classification layer to address the issue of averaging the output vectors in multi-class classification scenarios.

\end{itemize}
\section{Preliminary and Background}
\subsection{Neural Network Decision-making Process}
DNNs represent a cornerstone in the realm of machine learning. At their core, they are hierarchical models made up of multiple layers: typically an input layer, several hidden layers, and an output layer. The nodes of these layers are units known as neurons. Each neuron in a layer is connected to the neurons of the subsequent layer through connections weights and bias.

As input data is introduced into the network, it passes through each layer. The neurons compute a weighted sum of the inputs they receive, add a bias, and then apply a non-linear activation function like ReLU\cite{relu}. This processed output is then forwarded to the next layer. This cascade continues, layer by layer, until the data reaches the final layer, the output layer. For DNNs designed for classification tasks, the output layer often uses a softmax function. This function converts the raw outputs into probability distributions across the potential classes, providing a clear prediction.
Mathematically, the transformation occurring at hidden layer can be expressed as:

\begin{figure*}[tb]
	\centering
		\includegraphics[scale=0.55]{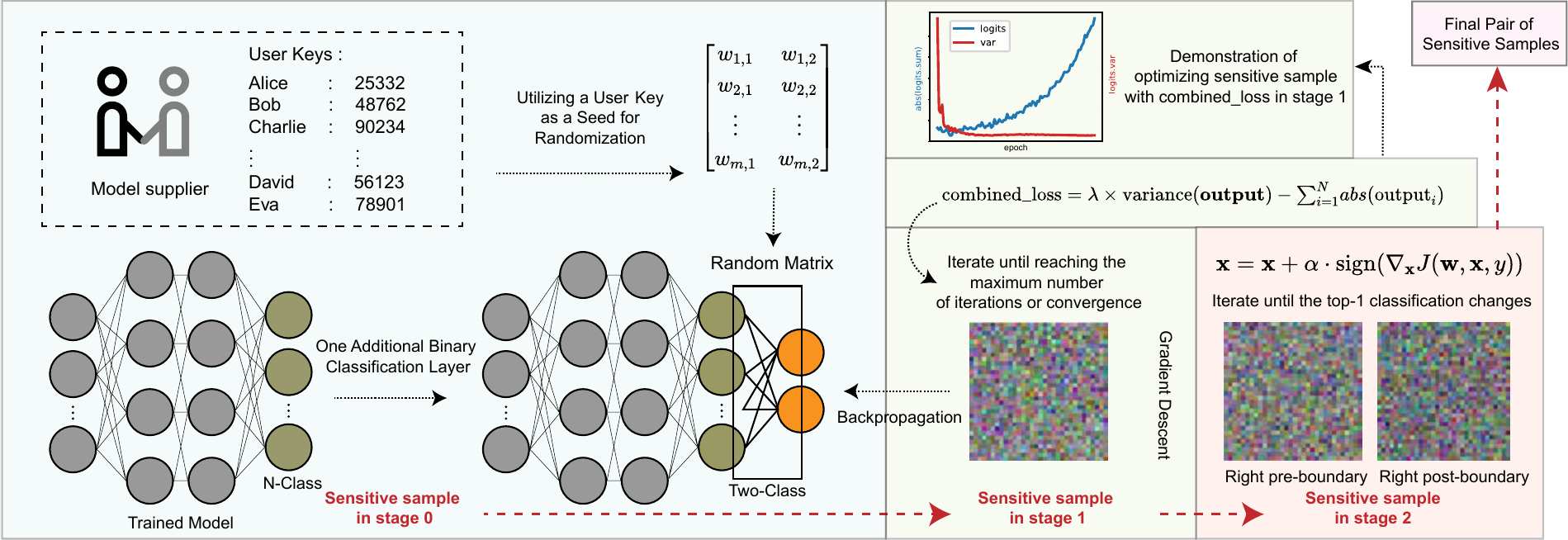}
	\caption{\small Overall framework for generating sensitive samples. In stage 0, the model supplier records users and their corresponding keys, and when users need to check the model, a weight matrix is generated using the corresponding keys to add one additional binary classification layer, facilitating user isolation for sensitive samples. In stage 1, under the existing the additional binary classification layer, we use a combined\_loss to optimize random pixels with the objective of bringing the sensitive samples close to the model boundary while maximizing neuron activation. Finally, in stage 2, using a method similar to adversarial attacks with a very small learning rate, we cross the model boundary and record the two sensitive samples before and after crossing to form a sample pair.}
	\label{figure2}
\end{figure*}

\begin{equation}
O_{L+1} = \phi_{L+1}(w_L \cdot O_L + b_L),
\end{equation}
where $O$ represents the activated value, and the subscript $L$ indicates which layer was activated, and $\phi$ represents the activation function (default in this article is ReLU). $w$ and $b$ correspond to weights and bias parameters. Due to the fact that neural networks are layer by layer, we can always use the output of the previous layer (close to the initial input) as the input of the next layer to complete the recursive process of the entire neural network. And final transformation can be represented as:

\begin{equation}
\small
\begin{aligned}
O_{L+1} = &\ \phi_{L+1}\left(w_{L} \times  \phi_{L}\left(w_{L-1} \times \ldots \right. \right. \\ 
&\ \left. \left. \times \phi_1\left(w_0 \times O_0 + b_0\right) \ldots + b_{L-1}\right) + b_{L}\right).
\end{aligned}
\end{equation}

\subsection{Model Boundary}

Consider a classification neural network designed to categorize incoming data. Its primary objective is to evaluate the input and return a confidence score for each potential category in the network. In a binary classification scenario, if the output scores for both classes are approximately 0.5, it suggests that the input sample is in close proximity to the model's decision boundary. Conversely, if the network's output strongly favors the confidence score of one class over the other, it indicates that the sample is situated far from the decision boundary.
Expanding this notion to multi-class scenarios and deeper network architectures introduces more complexity. Due to the involvement of multiple non-linear functions, the decision boundaries can take intricate forms. We can assume that these are all the results of mapping from high-dimensional space to low dimensional space. As the number of classification results continues to increase, it will become difficult for us to find a input data that is very close to all classification results (a similar classification probability). Here we provide a brief experiment to illustrate this issue. In different classification scenarios, using a step size of 1e-3 to approach the classification boundaries with 1000 rounds \cite{yin2023decision}, the absolute values of the coefficient of variation, representing the dispersion of the output logits, of each network are 0.07 for 10 classification networks, 4.46 for 43 classification networks, and 41.95 for 102 classification networks.

\section{Proposed method}

As shown in Figure \ref{figure2}, for models that require authorization for distribution, model suppliers establish a key repository for users. Using the corresponding keys, we construct a linear layer of size $N \times $2 for different users in stage 0. For this neural network, we employ a two-stage approach to generate highly sensitive for integrity-check samples. In stage 1, we iterate over the samples using the combine loss function until convergence is achieved or the maximum iteration steps are reached. The resulting samples from this stage then sent to the stage 2. In stage 2,  the samples utilize gradient ascent to modify the input samples with the aim of shifting the classification outcomes of the samples towards classification into another category. Stage 0, stage 1, and stage 2 will be further explained in the following subsection respectively.
\subsection{Adding One Additional Binary Classification Layer}

To differentiate users and mitigate the widespread leakage of sensitive samples, as well as to reduce the difficulty of achieving average output results, we employed one additional binary classification layer.
By recording the corresponding secret keys of users in a dictionary kept by the model supplier, weights for the linear layer converting N-class classification to binary classification are generated using these keys. This is used to add the corresponding binary linear layer during detection. This approach effectively reduces an N-class classification problem to a binary classification task, thereby simplifying the achievement of average classification outcomes. Consider a sample input into a binary classification network. If the output probabilities for each class are close to 50\%, this indicates uncertainty in the classification. Even minor changes to the original neural network can potentially alter the network's Top-1 output results.

\subsection{Training Loss Function}
Whether the modifications to the network are malicious or unintentional, we essentially impose a change of $\Delta w$ on the model parameters. This is reflected in the model and can be expressed as:
\begin{equation}
\small
\begin{aligned}
O_{i}(\omega+\Delta \omega, x)=
 O_{i}(\omega, x)+\frac{\partial O_{i}(\omega, x)}{\partial \omega} \Delta \omega+o\left(\|\Delta \omega\|_{2}^{2}\right).
\end{aligned}
\end{equation}

If we want to identify areas of the model that are particularly susceptible to changes, it is natural to look for regions where the value of $\frac{\partial O_{i}(\omega, x)}{\partial \omega}$ is as large as possible. And from the recursive formula of neural networks, we understand that the derivative of the network with respect to its parameters is influenced by the activation values of the current layer and the derivative values of the next layer. This relationship can be described through the formula:
\begin{equation}
\footnotesize
\begin{aligned}
\frac{\partial O_{L+2}}{\partial w_L}  = &\ \frac{\partial O_{L+2}}{\partial  O_{L+1}}\cdot\frac{\partial O_{L+1}}{\partial w_L} 
\\   =  &\  \phi^{\prime}(w_{L+1}\cdot O_{L+1}+b_{L+1})\cdot w_{L+1}\cdot\phi^{\prime}(w_L\cdot O_L+b_L)\cdot O_L.
\end{aligned}
\end{equation}

For the ReLU activation function, when the activation is greater than 0, the derivative is 1, and when it is less than 0, the derivative is 0. So the derivative value of the activation value for a certain layer parameter is closely related to the activation value of the layer where the parameter is located.

When considering all the layers that need activation, this can be written as:
\begin{equation}\label{eq5}
\small
\begin{aligned}
\frac{\partial O_L}{\partial w_{i-1}}=\left(\prod_{j=i}^{L-1}\frac{\partial O_{j+1}}{\partial O_j}\right)\cdot\phi^{\prime}(w_{i-1}\cdot O_{i-1}+b_{i-1})\cdot O_{i-1}.
\end{aligned}
\end{equation}

The gradient of each layer's parameters is influenced by the activation of the preceding layer. All of these influences ultimately converge on the activation of the final layer, which corresponds to the output logits. Consequently, we transition from the pursuit of maximum gradient to the pursuit of maximum activation, resulting in a significant reduction in computational resources required. We postulate that as we maximize the activation of the final layer to the greatest extent possible, the region in which the samples are situated becomes a highly responsive area within the model space, particularly sensitive to variations, as shown in Figure \ref{figure1}.

Based on Formula \ref{eq5}, we propose a methodology wherein simply maximizing the output of the last layer can activate more neurons. This activation further influences the gradient of related parameters, directing sensitive samples towards the regions of the model boundary with the highest susceptibility to change. To achieve proximity to the model boundary, we employ variance to average classification logits. Consequently, under this combined loss function, we have achieved results that are not only close to the model boundary but also positioned in areas where the boundary is most prone to variability, as shown in Formula \ref{eq6}.
\begin{equation}\label{eq6}
\footnotesize
\begin{aligned}
\mathrm{combined\_loss}=\lambda\times \mathrm{variance}(\mathrm{\mathbf{output}})-\sum_{i=1}^Nabs(\mathrm{output}_i)
\end{aligned}
\end{equation}

\subsection{Closely Approaching Model Boundaries}

Previous work utilized loss functions similar to  average classification outcome to get close to the model boundaries. However, this approach could lead to decreasing speed as one approaches the model boundaries (due to the decrease of loss), making it challenging to effectively control the approach rate. By employing gradient ascent, we can control the distance to the model boundary based on the optimization function for adversarial attacks, as shown in Formula \ref{eq7}:
\begin{equation}\label{eq7}
\begin{aligned}
\mathbf{x}=\mathbf{x}+\alpha \cdot \operatorname{sign}\left(\nabla_{\mathbf{x}} J(\mathbf{w}, \mathbf{x},y)\right),
\end{aligned}
\end{equation} 
where $\mathbf{x}$ represents the input to be optimized, $y$ is the original label, $\alpha$ represents the optimization step size, and $J(\mathbf{w}, \mathbf{x}, y)$ denotes the loss function of the model. This function quantifies the difference between the model's prediction for input $\mathbf{x}$ and the true label $y$, with $\mathbf{w}$ being the model's parameters. Given that the samples are already close to the sample boundary and exhibit high activation values, using smaller step sizes allows us to maintain these large activation values while also precisely approaching the model boundary. 

We record the sample information at two time points before and after the sample crosses the model boundary, and combine it into a sample pair, completing the construction of sensitive sample pairs, as shown in the selected samples in Figure \ref{figure1}.

\section{Experiment}

In this section, we evaluate the proposed fragile model watermark based on sensitive samples in terms of sensitivity and generation efficiency. For experiment assessment, we embed backdoors into the model, perform fine-tuning with a small learning rate, pruning, and 8-bit quantization, and compare these with other sensitive sample methods. The experiments were carried out on a system configured with 1 NVIDIA RTX3090 GPU and 2 Intel Xeon(R) Silver 4210R CPUs. All input tensors are normalized to [0,1] like other methods. The information about the models and dataset settings used in our study, implemented using PyTorch, can be found in Table \ref{Table:datasets}.

\begin{table}[ht]
\footnotesize
\centering
\caption{Datasets and models.}
\label{Table:datasets}
\begin{tabular}{l||cccc}
\toprule
\textbf{Dataset}    & \textbf{Resize}     & \textbf{Classes} & \textbf{Model} &\textbf{Accuracy(\%)}   \\
\midrule
Cifar10             &  3$\times$32$\times$32          & 10               &   ResNet18  &91.26 \\
GTSRB               &  3$\times$40$\times$40          & 43               &   VGG16        &96.70\\
Flowers102          &  3$\times$128$\times$128         & 102              &  ResNet152 &98.41\\
\bottomrule
\end{tabular}
\end{table}

\subsection{Effectiveness in Sensitivity}

In our study, we define successful detection as a change in the Top-1 label output by the neural network. Throughout subsequent model adjustments, we consistently use 150 as the numerical base. For our method, this necessitates the use of 150 pairs of sensitive samples. To ensure that any performance improvement is not merely due to a greater number of samples required by our method compared to others, we form sample pairs for other methods, where a change in the Top-1 label in either sample indicates successful detection. In our method, the learning rate is set at 1e-3 during stage 1, while in stage 2, it is 1e-4 for the VGG16 model and 1e-6 for others. For the AID \cite{aid} method, a learning rate of 1e-1 is used on VGG16, while a rate of 1e-3 is employed for other cases and for DBI \cite{yin2023decision}. The `Validset' in the table refers to samples randomly selected from the validation set, with a total number of 150. Our approach, along with other methodologies, employs a maximum iteration count of 10,000.

\begin{table}[htb]
\footnotesize
\centering
\caption{\small Detection success rate (\%) of sensitive samples post-model backdoor attack. The numbers to the left of the slash represent the detection probability for a single sample, while those on the right represent the detection probability for a pair of samples randomly selected.}
\label{Table:backdoor}
\begin{tabular}{l||ccc|c}
\toprule
\textbf{Dataset} & \textbf{Validset} & \textbf{DBI \cite{yin2023decision}} & \textbf{AID \cite{aid}} & \textbf{Ours} \\
\midrule
Cifar10          & 0.86        & 62.46 / 86.13 & 61.59 / 85.40 & 99.59       \\
GTSRB            & 0.00        & 46.47 / 73.66 & 47.80 / 74.93 & 92.61         \\
Flowers102       & 3.20        & 27.47 / 47.60 & 29.27 / 48.47 & 99.20         \\
\bottomrule
\end{tabular}
\end{table}

In our approach to backdoor attacks, we chose to implant 10 samples with markers and designated labels into the model, training them alongside normal samples (all using a learning rate of 1e-5) until the network completely recognized the intended backdoor as per the specified label. The numbers left of the slash in our experimental results represent the probability of a test sample detecting a change in the model, while the numbers right of the slash indicate the probability of either of the two samples detecting a change, as shown in Table \ref{Table:backdoor}. It is evident that our method significantly outperforms others in identifying implanted backdoors. 

Since the backdoor implantation process involves fine-tuning with a small learning rate, we further tested our method's detection rate of model modifications under extremely small learning rates, as shown in Table \ref{Table:finetune}. Remarkably, even under the rare condition of a 1e-9 learning rate, our method's detection rate on ResNet18 and ResNet152 remains over 74\%. Similarly, we observed that at this learning rate, the validsets recognition rate dropped to zero, indicating almost no change in the model in relation to the validset and minimal variation in the model boundaries compared to natural samples.

\begin{table}[ht]
\footnotesize

\centering
\caption{Detection Success Rate (\%) of Sensitive Samples Post-Model Fine-Tuning.}
\label{Table:finetune}
\begin{tabular}{p{0.8cm}||cccc|c}
\toprule
\textbf{Dataset} & \textbf{Lr} & \textbf{Validset} & \textbf{DBI \cite{yin2023decision}} & \textbf{AID \cite{aid}} & \textbf{Ours} \\
\midrule
\multirow{6}{*}{Cifar10}
        & 1e-10 & 0.00 & 19.69 / 38.01 & 18.21 / 48.36 & 49.71 \\
        & 1e-09 & 0.00 & 22.42 / 39.82 & 20.70 / 49.02 & 77.56 \\
        & 1e-08 & 0.01 & 31.92 / 49.47 & 25.19 / 51.13 & 99.56 \\
        & 1e-07 & 5.46 & 30.67 / 49.34 & 25.33 / 51.88 & 100 \\
        & 1e-06 & 0.00 & 31.33 / 49.85 & 25.44 / 53.14 & 100 \\
        & 1e-05 & 12.68 & 31.24 / 51.66 & 25.38 / 50.74 & 100 \\
        \midrule
\multirow{6}{*}{GTSRB}
        & 1e-08 & 0.00 & 47.98 / 71.75 & 53.3 / 75.64 & 51.43 \\
        & 1e-07 & 0.00 & 47.34 / 75.33 & 51.40 / 74.63 & 88.42 \\
        & 1e-06 & 0.00 & 47.32 / 75.31 & 53.42 / 72.41 & 93.72 \\
        & 1e-05 & 0.00 & 47.33 / 76.04 & 51.02 / 72.17 & 97.59 \\
        & 1e-04 & 4.11 & 47.33 / 75.32 & 42.71 / 72.02 & 99.99 \\
        & 1e-03 & 39.25 & 47.26 / 75.33 & 43.98 / 72.19 & 100 \\
        \midrule
\multirow{6}{*}{\parbox{2cm}{Flowers\\ \hspace{5mm}102}}
           & 1e-10 & 0.00 & 3.78 / 6.04 & 3.43 / 8.52 & 68.33 \\
           & 1e-09 & 0.00 & 3.83 / 6.28 & 3.78 / 8.81 & 74.01 \\
           & 1e-08 & 0.01 & 6.67 / 17.12 & 8.86 / 15.39 & 91.82 \\
           & 1e-07 & 4.77 & 18.22 / 32.91 & 20.99 / 36.06 & 98.51 \\
           & 1e-06 & 8.63 & 15.42 / 33.05 & 26.65 / 36.78 & 99.85 \\
           & 1e-05 & 11.95 & 13.81 / 33.56 & 27.36 / 36.28 & 99.99 \\
\bottomrule
\end{tabular}
\end{table}

\begin{figure}[htb]
	\centering
		\includegraphics[scale=0.35]{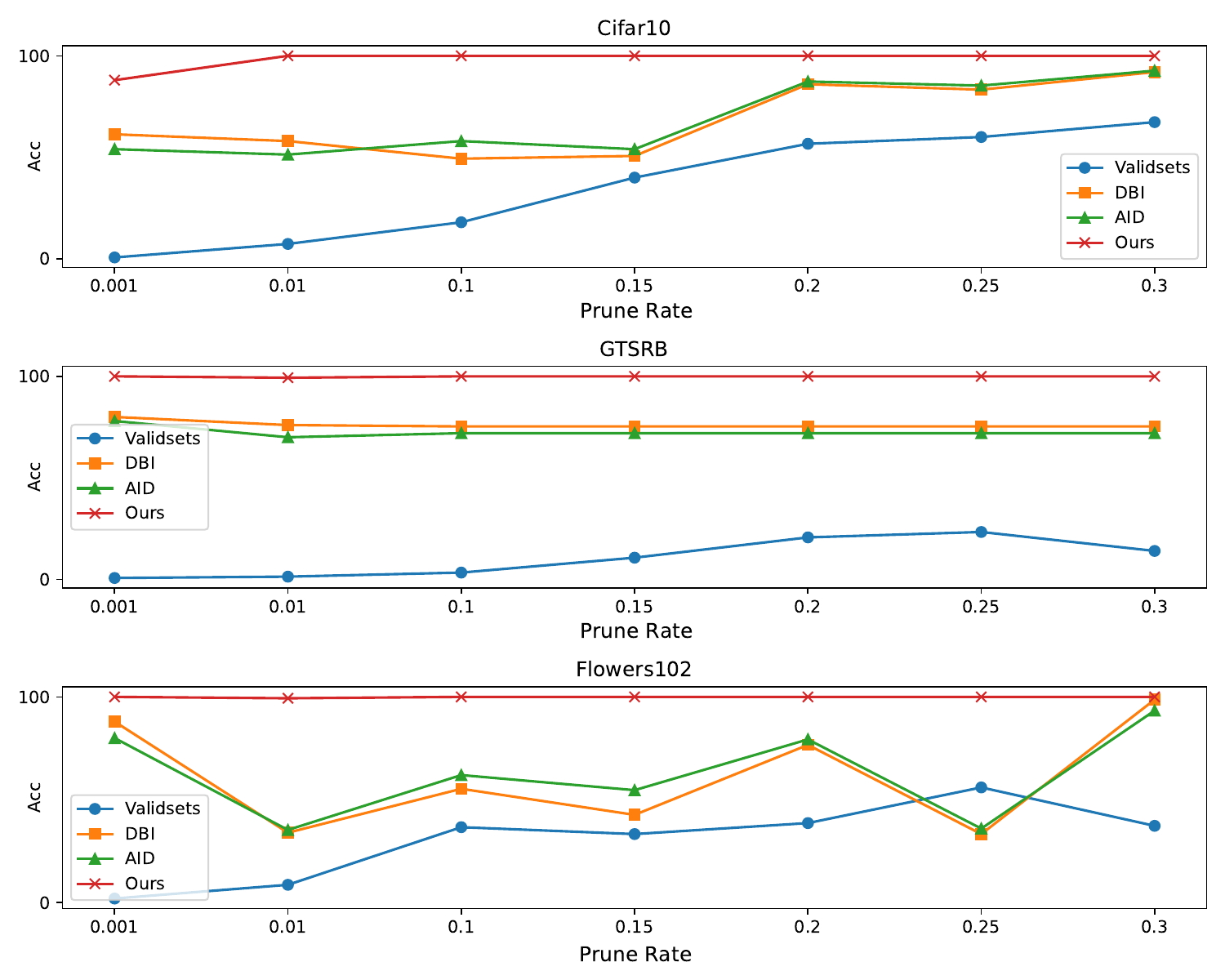}
	\caption{The Success Rate (\%) of Detecting Sensitive Samples After
Prune the Models.}
	\label{F.Prune}
\end{figure}

Our study observes a similar effect in model pruning, as illustrated in Figure \ref{F.Prune}. Even with 0.1\% pruning, our method maintains high detection efficiency. However, under quantization, as Table \ref{T.quantized} shows, the improvement in detection accuracy is not significant in the setting of sample pairs. Still, our method outperforms others in single-sample scenarios. We speculate this is because, under sample pair conditions, other sensitive samples were randomly combined pairs, whose accuracy improvements relied on the independence between pairs (a setup in our experiment, not an inherent requirement of the other methods). In contrast, our sample pairs, derived from adversarial attacks. Given the reality that we cannot a prior know the treatments a model we aim to detect may have undergone, our method's high detection rate conditions still proves its effectiveness to a certain extent.

\begin{table}[ht]
\footnotesize
\centering
\caption{Detection Success Rate (\%) of Sensitive Samples Post-Model 8-bit Quantize.}
\label{T.quantized}
\setlength\tabcolsep{3pt}
\begin{tabular}{l||ccc|cc}
\toprule
\textbf{Dataset} & \textbf{Validsets} & \textbf{DBI \cite{yin2023decision}} & \textbf{AID \cite{aid}} & \textbf{Unpaired} & \textbf{Paired} \\
\midrule
Cifar10     & 0.67  & 28.67 / 54.67 & 34.00 / 64.00  & 41.33 & 44.67 \\
GTSRB       & 0.00  & 59.33 / 82.00    & 58.67 / 83.33 & 62.00 & 79.33 \\
Flowers102  & 0.67  & 33.33 / 64.00    & 33.33 / 50.67 & 47.33 & 49.33 \\
\bottomrule

\end{tabular}
\textit{Note: 'Unpaired' indicates that only one sample from the sample pair is used for detection in our method, while 'Paired' represents the normal sample pairs.}
\end{table}

\subsection{Efficiency in generating sensitive samples}
To evaluate efficiency we focus on comparing the core optimization functions of our respective schemes. In order to compare the same effect, we ``binarize" all models and optimize sensitive samples. Due to the existence of stage 2 in our scheme, while stage 1 is a similar step to other schemes, the time shown in Table \ref{T.efficiency} is the time occupied in stage 1, However, this does not affect the overall operation of our scheme. For the previous experiment, when our stage 1 is close enough to the model boundary after 10000 iterations, we only need less than 60 seconds of iteration on the above dataset and model to obtain two samples on both sides of the model boundary. This means that in a very short amount of time, we have also managed to double the number of samples, forming pairs of samples. The core difference among the three schemes lies in whether there are other tricks that make the sensitive samples more sensitive to model changes, while the sensitive samples are close to the model boundary. Since DBI \cite{yin2023decision} is only a simple approach to the classification boundary, it is natural to use the least amount of resources, however, due to the lack of coverage or activation of neurons, the detection success rate may significantly decrease when using lower learning rates on the model (such as 1e-9, 1e-10). On the contrary, AID \cite{aid} traverses all neurons, which makes it difficult to determine hyperparameters and reduce iteration speed when we need to count more and more neurons when traversing larger models. Of course, we need to affirm the effectiveness of its scheme in terms of detection success rate. As the learning rate decreases, the detection success rate remains at its original performance. In stage 1, we reconsidered backpropagation to avoid using the method of traversing all neurons and only focusing on the final output logits. Based on this, our scheme's sensitive sample iteration speed is already very close to only averaging the output results, while still exhibiting a certain degree of robustness at low learning rates.

\begin{table}[ht]
\scriptsize
\centering
\caption{\small Time required for 1000 iteration rounds per Method's Core Optimization Function. As DBI primarily focuses on approaches near the boundary, it naturally takes the least amount of time. In contrast, our method, which involves transforming the gradient problem into a final logits issue, requires only slightly more time than DBI, significantly outperforming the AID method. }
\label{T.efficiency}
\begin{tabular}{l|c|c|c|c|c}
\toprule
\textbf{Model} & \textbf{Num.} & \textbf{Params(M)} & \textbf{DBI(s)} & \textbf{AID(s)} & \textbf{Ours(s)} \\
\midrule
\multirow{2}{*}{ResNet18}   & 150 & \multirow{2}{*}{11.69} & 13.92 & 24.48 & 14.11 \\
           & 300 &       & 15.68 & 28.68 & 15.96 \\ \midrule
\multirow{2}{*}{VGG16}      & 150 & \multirow{2}{*}{138.34} & 24.83 & 40.98 & 26.42 \\
           & 300 &       & 43.73 & 66.91 & 44.46 \\ \midrule
\multirow{2}{*}{ResNet152}  & 150 & \multirow{2}{*}{60.19} & 183.32 & 429.93 & 184.54 \\
           & 300 &       & 401.56 & 765.31 & 408.38 \\
\bottomrule
\end{tabular}
\textit{\\ Note: "Num." represents the number of samples. }
\end{table}
\section{Conclusion}
In this paper, we construct sensitive samples that achieve high sensitivity detection of minor alterations to the model, while also ensuring efficient generation. We utilize one additional binary classification layer to simplify the process of closely approaching the model boundary and transform the gradient problem into an activation problem, thereby reducing the resource consumption associated with gradient calculation or neuron traversal. Specifically, we consider both proximity to the classification boundary and the ease of change within the boundary region. We achieve this by maximizing logits on the basis of average classification outcomes and then employing very small step sizes to closely converge towards the model boundary. This approach ensures that the samples are exceedingly close to the model boundary, which remains highly susceptible to change at this juncture. Our experiments demonstrate the effectiveness of our approach, showcasing its superior performance across multiple metrics.


\bibliographystyle{IEEEbib}
\small
\bibliography{icme2023template}

\end{document}